\documentclass[10pt,twocolumn]{article}

\usepackage[utf8]{inputenc}
\usepackage[T1]{fontenc}
\usepackage{graphicx} % For figures
\usepackage{amsmath,amssymb} % Standard math
\usepackage{geometry}
\usepackage{setspace}
\usepackage{lineno} % Line numbers (optional but common in submission)
\usepackage{cite} % Numeric citations
\usepackage{hyperref} % For clickable links (safe for submission)
\usepackage{booktabs}
\usepackage{booktabs}
\usepackage[table]{xcolor}
\usepackage{xcolor}
\usepackage{svg}

\usepackage{xspace}
\usepackage{adjustbox}
\usepackage{textcomp}
\usepackage{tablefootnote}
\usepackage{threeparttable}
\usepackage{authblk}

\definecolor{lightblue}{rgb}{0.8,0.85,1}

\makeatletter
\@ifpackageloaded{xspace}{
  \xspaceaddexceptions{"'}
}{}
\makeatother
\newcommand*{\aefm}{AEFM\xspace}

\title{Adaptive Transition State Refinement with\\ Learned Equilibrium Flows}

\author[1,2]{Samir Darouich}
\author[2]{Vinh Tong}
\author[2]{Tanja Bien}
\author[1]{Johannes Kästner}
\author[2]{Mathias Niepert}
\affil[1]{Department of Chemistry, University of Stuttgart, Germany}
\affil[2]{Department of Artificial Intelligence, University of Stuttgart, Germany}
\date{}

\begin{document}

\maketitle

\begin{abstract}

Identifying transition states (TSs), the high-energy configurations that molecules pass through during chemical reactions, is essential for understanding and designing chemical processes. However, accurately and efficiently identifying these states remains one of the most challenging problems in computational chemistry. In this work, we introduce a new generative AI approach that improves the quality of initial guesses for TS structures. Our method can be combined with a variety of existing techniques, including both machine learning models and fast, approximate quantum methods, to refine their predictions and bring them closer to chemically accurate results. Applied to TS guesses from a state-of-the-art machine learning model, our approach reduces the median structural error to just 0.088~\AA\ and lowers the median absolute error in reaction barrier heights to 0.79~kcal~mol$^{-1}$. When starting from a widely used tight-binding approximation, it increases the success rate of locating valid TSs by 41\% and speeds up high-level quantum optimization by a factor of three. By making TS searches more accurate, robust, and efficient, this method could accelerate reaction mechanism discovery and support the development of new materials, catalysts, and pharmaceuticals.

% Transition states (TS) are essential for modeling and understanding chemical reaction mechanisms. Despite their importance, locating TS remains one of the most significant bottlenecks in computational chemistry, a challenge that has recently seen progress through machine learning approaches. In this work, we introduce \aefm (Adaptive Equilibrium Flow Matching), an algorithmic framework designed to enhance the chemical accuracy of initial TS guesses. \aefm is model-agnostic and can be seamlessly integrated with low-fidelity methods such as ML-based predictors or tight-binding approximations. Applied to TS guesses from the state-of-the-art generative model React-OT, \aefm reduces the median structural root mean square deviation to 0.088 Å and achieves a median absolute error in barrier heights of 0.793 kcal mol$^{-1}$. When starting from GFN2-xTB estimates, \aefm increases the rate of valid TSs by 41\% and accelerates DFT-based TS optimization by a factor of three. We anticipate that \aefm will advance the speed and reliability of reaction mechanism elucidation and thereby contribute to the efficient design and optimization of chemical processes.

\end{abstract}
\section{Introduction}\label{sec1}

The transition state (TS) plays a central role in elucidating reaction mechanisms and understanding the microkinetic behavior of chemical processes\cite{truhlar_current_1996, peng2016computing, dewyer_methods_2018, von_lilienfeld_exploring_2020, unsleber_exploration_2020, nandy_computational_2021,jorner_organic_2021}. 
A detailed knowledge of the underlying kinetics enables the rational design of catalysts, synthetic routes, and functional materials, driving progress toward more efficient, sustainable, and innovative chemical processes\cite{taylor2023brief,chacko2024interconnected}. Computationally, a TS corresponds to a first-order saddle point on the potential energy surface (PES). Algorithms for locating TSs fall into two broad categories, single-ended\cite{ banerjee_search_1985, baker_algorithm_1986, henkelman_dimer_1999} and double-ended methods\cite{jonsson1998nudged,henkelman_climbing_2000, peters_growing_2004}. Single-ended methods refine an initial 3D structure using gradient and sometimes Hessian information, while double-ended approaches construct a continuous path between reactant and product geometries to locate the TS along this path. However, when based on high-level electronic structure methods such as density functional theory (DFT)\cite{mardirossian2017thirty}, these algorithms become prohibitively expensive, posing a significant bottleneck in reaction mechanism discovery. 

To overcome this limitation, machine learning (ML) has emerged as a promising direction. Surrogate models, such as Gaussian process regressions\cite{pozun_optimizing_2012,koistinen_minimum_2016,koistinen_nudged_2017,denzel_gaussian_2018,denzel_gaussian_2019, garrido_torres_low-scaling_2019, heinen_transition_2022} or machine-learned interatomic potentials\cite{peterson_acceleration_2016,schreiner_neuralnebneural_2022, zhang2024exploring, wander_cattsunami_2024, yuan_analytical_2024, zhao_harnessing_2025}, can approximate the PES, significantly accelerating TS searches when coupled with traditional optimization schemes. However, these approaches require high-quality non-equilibrium data, particularly around the TS region, which limits their scalability\cite{yuan_analytical_2024}. Beyond surrogate-assisted optimization, other deep learning approaches aim to directly predict the transition state structure\cite{pattanaik_generating_2020, jackson_tsnet_2021, zhang_deep_2021, choi_prediction_2023}. Many of these methods predict the TS distance matrix and then convert it into 3D coordinates. More recently, generative models have reframed TS prediction as a distribution learning problem, aiming to learn the distribution of TS geometries conditioned on given reactant and product structures\cite{makos_generative_2021, duan_accurate_2023, kim_diffusion-based_2024, galustian_goflow_2025, duan_optimal_2025, hayashi_generative_2025}. For instance, ReactDiff\cite{duan_accurate_2023} models the joint distribution of reactant, TS, and product using denoising diffusion and inpainting to sample plausible TS candidates. Its successor, React-OT\cite{duan_optimal_2025}, leverages flow matching (FM)\cite{lipman_flow_2022, liu_flow_2022} and optimal transport to improve generation accuracy and efficiency. Other models bypass the need for 3D input entirely by generating TS geometries directly from 2D molecular graphs\cite{ kim_diffusion-based_2024, galustian_goflow_2025}. While generative models are promising, they can struggle to resolve fine-grained geometric details, sometimes producing unphysical features such as atomic collisions or distorted bond lengths\cite{peng_moldiff_2023,williams_physics-informed_2024, vost_improving_2025,wohlwend_boltz-1_2025, galustian_goflow_2025}. In contrast, TS guesses from approximate quantum chemical methods, like tight-binding, tend to be physically plausible, but can systematically deviate from DFT-level structures\cite{rasmussen2020fast}. In both scenarios, the predicted TS structures function as low-fidelity approximations that, while providing valuable initial estimates for reaction exploration, may require further refinement to achieve the accuracy needed for quantitatively reliable kinetic analysis.

To address this gap, we introduce Adaptive Equilibrium Flow Matching (\aefm), a structure-only refinement method that transforms low-fidelity TS guesses, regardless of their origin, into high-accuracy transition state geometries without requiring any energy or gradient information. \aefm learns to invert noise-injected perturbations of reference TS structures using a novel time-independent form of variational flow matching (VFM)\cite{amini2024variational}. The model operates by predicting integration steps that iteratively refine the structure, converging toward a fixed-point solution. By additionally respecting the symmetry inherent in molecular structures, \aefm introduces a $\mathrm{SE}(3)$-equivariant method that facilitates robust inference, adaptable to the quality of the initial TS structure. To further improve the chemical realism of refined structures, we incorporate a physics-inspired bond-based loss that guides the model toward physically plausible geometries. \aefm is particularly suited for high-throughput settings, where efficient and reliable refinement is essential to handle large numbers of candidates. Additionally, it benefits in-depth mechanistic studies by reducing the need for costly TS optimization steps. When used in conjunction with React-OT, a state-of-the-art ML-based model, \aefm reduces the median root-mean-square deviation (RMSD) of predicted TS structures to 0.088~\AA\ and achieves a median absolute error in barrier heights of just 0.793~kcal~mol$^{-1}$, a 27\% improvement over React-OT alone. Incorporating a physics-inspired bond-length loss further enhances structural realism, with the bonded distance distribution of \aefm-refined samples aligning 35\% more closely to the ground truth distribution from the Transition1x dataset\cite{schreiner_transition1x_2022}. \aefm also boosts the chemical validity rate of GFN2-xTB-generated\cite{Bannwarth2021} TSs by 41\%, making the combined approach a practical solution for rapid and accurate TS discovery.
\section{Results}\label{sec2}

\subsection*{Method overview}\label{subsec2_1}

\begin{figure*}[ht]
\centering
\includegraphics[width=0.9\textwidth]{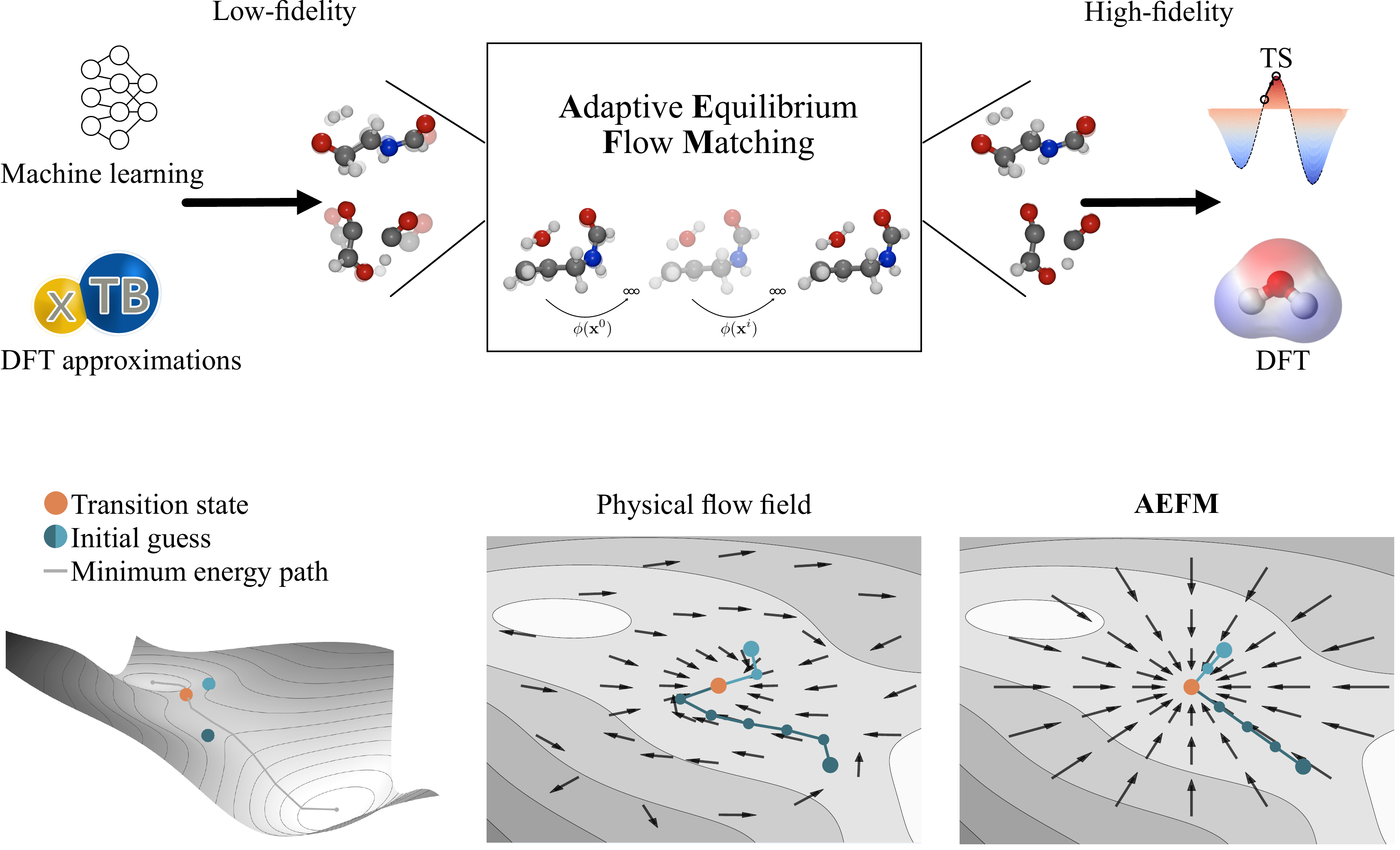}
\caption{
\textbf{\aefm pipeline for TS structure refinement.}
\textbf{a} The input consists of low-fidelity TS samples, which may originate from various sources such as ML models or tight-binding approximations. These inputs are iteratively refined to produce high-fidelity, chemically valid TS geometries near the DFT level.
\textbf{b} Comparison between actual physical flow and the one learned by \aefm on the Müller–Brown potential energy surface. Integrating the physical flow field requires multiple function evaluations, which can become computationally expensive with methods such as DFT. In contrast, \aefm learns a much simpler representation that captures the essential structure while requiring significantly fewer and more efficient evaluations.
}\label{fig:method}
\end{figure*}

\aefm builds on the principle of FM, which learns to transform samples from one distribution into another. The transformation is done by learning a time-dependent vector field that transports samples from the prior distribution to the target distribution along a predefined probability path. In our case, as illustrated in Figure~\ref{fig:method}, the goal is to refine low-fidelity TS structures, such as those predicted by ML models or semiempirical methods, into high-quality TS geometries. To train \aefm, we start from accurate TS structures taken from a reference dataset, which define the target distribution. To reflect the kind of inaccuracy expected from low-fidelity initial guesses, we perturb the reference structures with noise proportional to the prior method's typical error. This scaling allows \aefm to adapt to the error magnitude of the input source, enabling it to generalize across different prior methods. \aefm then learns a continuous transformation, guided by optimal transport, that maps these noisy inputs back to their original high-fidelity TS geometries. At inference time, however, the quality of low-fidelity TS samples varies, as some may already lie close to the desired distribution, while others deviate significantly. A standard FM inference scheme would apply a uniform integration of the velocity field across all inputs, which can lead to under- or overshooting depending on the initial error. To address this, we instead train \aefm without a time-dependent formulation, resulting in a time-independent equilibrium flow field. This field consistently points toward the high-fidelity structures, enabling fixed-point inference that iteratively pulls each sample toward its refined geometry, regardless of its initial deviation. The number of refinement steps adapts dynamically to the quality of the input, allowing the model to allocate computational effort where it is most needed. To respect molecular symmetries, such as rotation, translation, and atom index permutation, we employ the SE(3)-equivariant LEFTNet\cite{du2023new} architecture as the backbone of our model. In addition, we propose to incorporate domain knowledge through a bond-loss term, which encourages physically plausible predictions by penalizing unrealistic bond configurations. This explicit inclusion of physics-based constraints guides the model toward more chemically meaningful outputs.

\aefm introduces several methodological innovations to enable efficient and accurate refinement of TS structures. Unlike standard FM, which relies on a time-dependent vector field and fixed integration schedules, \aefm learns a time-independent equilibrium flow field that supports adaptive fixed-point inference. To promote chemically realistic outputs, a physics-inspired bond-length loss that penalizes implausible bond distortions is incorporated. Together, these design choices enable a structure-only training and inference pipeline, eliminating the need for potential energy surface evaluations or gradient computations. As a result, \aefm achieves up to a 40\% reduction in mean barrier height errors within just four inference steps, typically completing in less than a second. When quantum mechanical optimization is still required, \aefm-refined structures lead to a threefold reduction in computation time compared to unrefined inputs, substantially lowering computational cost while improving stability.

\subsection*{Refining TS structures across fidelity scales}\label{subsec2_2}

To evaluate \aefm, we use the Transition1x dataset\cite{schreiner_transition1x_2022}, which contains climbing-image nudged elastic band (CI-NEB)\cite{henkelman_climbing_2000} calculations performed with DFT ($\omega$B97x/6-31G(d)\cite{ditchfield1971self,chai2008systematic}) for 10,073 organic reactions encompassing diverse reaction types. These reactions were sampled from an enumeration of 1,154 reactants in the GDB7 dataset\cite{grambow2020reactants}, which includes molecules with up to 7 heavy atoms (C, N, and O) and a total of 23 atoms. We adopt the same random split as Duan et al.\cite{duan_accurate_2023}, using 9,000 reactions for training and 1,073 for testing.

\aefm is applied to refine prior low-fidelity TS structures toward valid TS geometries at the target level of theory. To assess the quality of the refined structures, we evaluate both the RMSD of atomic positions and the absolute error in the reaction barrier.

\begin{figure*}[ht]
\centering
\includegraphics[width=0.9\textwidth]{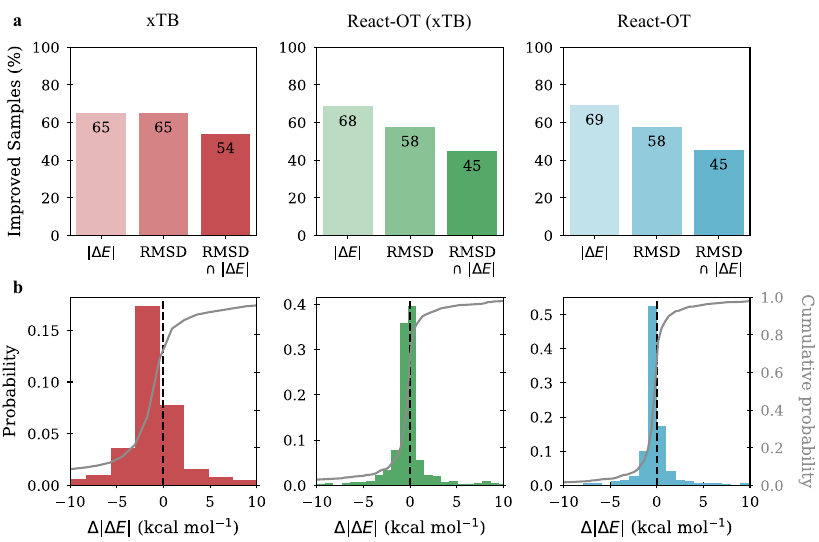}
\caption{
\textbf{Performance summary of \aefm across diverse low-fidelity sources.}
\textbf{a} Percentage of test samples showing improvement in energy difference $|\Delta E|$ relative to the reference TS (irrespective of RMSD), in RMSD (irrespective of energy), and in both RMSD and energy difference (RMSD $\cap$ $|\Delta E|$).
\textbf{b} Histogram (colored, left y-axis) and cumulative distribution (grey, right y-axis) of the change in energy difference between the low-fidelity and \aefm fine-tuned samples, measured relative to the reference TS. Negative $|\Delta E|$ values indicate that the refined samples are energetically improved.}\label{fig:stats}
\end{figure*}

\begin{table*}[ht]
\centering
\caption{Performance of \aefm refinement. Structural and energetic errors of various low-fidelity TS guesses before and after refinement with \aefm. The refinement consistently reduces both mean and median deviations relative to the reference TS structures. In addition, average inference times per sample are reported, showing that \aefm introduces only negligible computational overhead.
}
\vspace{0.5em}
\label{tab:performance}
\begin{threeparttable}
\begin{adjustbox}{max width=\linewidth}
\begin{tabular}{@{\extracolsep\fill}lllllc}
\toprule
Approach & \multicolumn{2}{c}{RMSD (\AA)} & \multicolumn{2}{c}{$|\Delta E_{\text{TS}}|$ (kcal mol$^{-1}$)} & Inference (s) \\
\cmidrule(lr){2-3} \cmidrule(lr){4-5}
& Mean & Median & Mean & Median \\
\midrule
xTB CI-NEB  & 0.312  & 0.179 & 10.426 & 2.673 & 9.23 \\
\rowcolor{lightblue}
xTB CI-NEB + \aefm & 0.250 ($\downarrow$20\%) & 0.119 ($\downarrow$34\%) & 6.204 ($\downarrow$40\%) & 1.090 ($\downarrow$59\%)& +0.24 \\
React-OT (xTB)  & 0.211  & 0.108  & 4.697 & 1.186 & 0.14 \\
\rowcolor{lightblue}
React-OT (xTB) + \aefm & 0.214 ($\uparrow$1\%) & 0.102 ($\downarrow$6\%) & 4.153 ($\downarrow$12\%) & 0.824 ($\downarrow$31\%) & +0.12 \\
React-OT  & 0.183  & 0.092 & 3.405  & 1.092 & 0.14 \\
\rowcolor{lightblue}
React-OT + \aefm & 0.188 ($\uparrow$3\%) & 0.088 ($\downarrow$4\%) & 3.341 ($\downarrow$2\%) & 0.793 ($\downarrow$27\%) & +0.13 \\
\rowcolor{lightblue}
React-OT + \aefm\tnote{a} & 0.176 ($\downarrow$4\%) & 0.086 ($\downarrow$7\%) & 3.158 ($\downarrow$7\%) & 0.790 ($\downarrow$27\%) & +0.13\\
\bottomrule
\end{tabular}
\end{adjustbox}
\begin{tablenotes}
\footnotesize
\item[a] For 26 reactions, a different intended TS was selected if the RMSD between the low-fidelity sample and this alternative TS was at least 30\% lower than the RMSD to the originally intended TS.
\end{tablenotes}
\end{threeparttable}
\end{table*}

To assess the effectiveness of \aefm, we consider React-OT\cite{duan_optimal_2025} as the first low-fidelity source, a state-of-the-art generative model for TS prediction. React-OT achieves remarkable accuracy, producing samples with a mean RMSD of 0.18~\AA\ and a median absolute error in barrier height of 1.092~kcal~mol$^{-1}$. Applying \aefm to refine the React-OT samples yields a 27\% improvement in the median barrier height error, requiring only 2 model calls in median and approximately 0.13 seconds per refinement on an Nvidia A40 GPU. Consequently, 69\% of the TSs had a more accurate barrier height, achieving a median absolute error of 0.793~kcal~mol$^{-1}$. 

As a second low-fidelity source, we consider GFN2-xTB\cite{Bannwarth2021}, a tight-binding approximation that is commonly used as a starting point for elucidating reaction mechanisms. Tight-binding methods are approximately three orders of magnitude faster than DFT, enabling high-throughput reaction scans that would be otherwise computationally prohibitive. For the 1,073 test reactions, reactant and product geometries were first relaxed, followed by CI-NEB calculations using GFN2-xTB. Of these, 945 calculations converged successfully, yielding samples with a mean RMSD of 0.31~\AA\ and a median absolute error in barrier height of 2.673~kcal~mol$^{-1}$. Applying \aefm improves the median absolute error in barrier height by 59\%, reducing it to 1.090~kcal~mol$^{-1}$, while requiring only a median of 4 model calls. To contextualize this improvement, in mikrokinetic modeling, an error of one order of magnitude change in reaction rate is considered as chemical accuracy, corresponding to  1.58~kcal~mol$^{-1}$ error in barrier height at 70$^\circ$C\cite{bremond2022tackling,fu2022meta}. Analyzing the chemical accuracy of samples reveals that only 25\% of the original GFN2-xTB-generated structures meet this threshold, whereas \aefm refinement increases this accuracy rate to 57\%.

To reduce the computational cost of generating DFT-quality reactant and product structures, we follow Duan et al.~\cite{duan_optimal_2025} and employ React-OT directly on xTB-optimized geometries. This approach enables rapid TS generation without requiring expensive DFT-level optimization of endpoints. React-OT can be reliably applied to xTB-level structures, yielding a mean RMSD of 0.21~\AA\ and a median absolute error in barrier height of 1.186~kcal~mol$^{-1}$. Building on this, we apply \aefm to refine the resulting TS guesses further, reducing the median absolute error by an additional 31\% with only a median of two model evaluations. The results of \aefm applied to each low-fidelity method are summarized in Table~\ref{tab:performance}.

\subsection*{Understanding refinement dynamics}\label{subsec2_3}

\begin{figure*}[ht]
\centering
\includegraphics[width=0.9\textwidth]{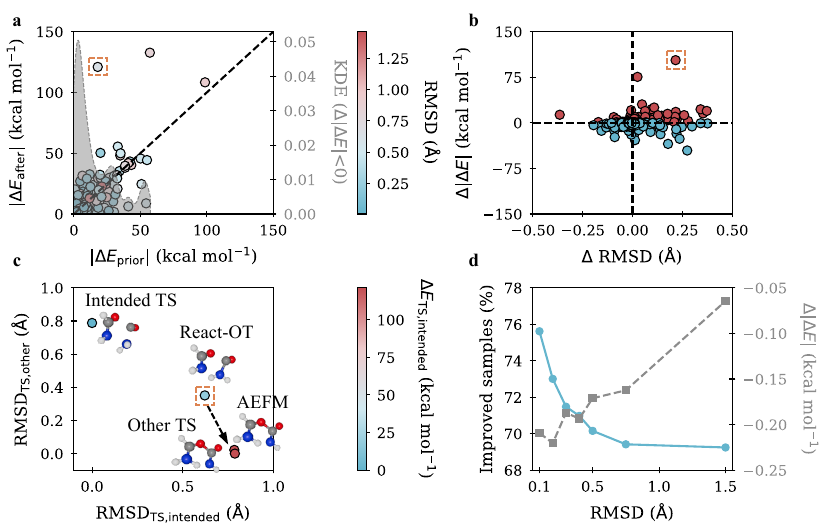}
\caption{
\textbf{Relationship between energetic and structural changes in \aefm\ refinements, with a focus on outliers and correlation trends.}
\textbf{a} Energetic differences of \aefm-refined structures versus initial React-OT predictions on the left y-axis. Points below the diagonal line indicate improved agreement with the reference TS, while points above reflect increased deviation. On the right y-axis, the KDE of improvement weighted by the improvement magnitude is shown. Additionally, an outlier (top left) shows a nearly sixfold increase in error after fine-tuning. 
\textbf{b} Energetic vs. geometric changes resulting from the application of \aefm. The bottom-left quadrant indicates improvements in both structural and energetic similarity, while the bottom-right quadrant reflects improved energy alignment accompanied by reduced structural similarity. Blue points indicate an energetic improvement, while red points correspond to increased dissimilarity.
\textbf{c} Structural analysis of the outlier. The x-axis shows RMSD to the intended TS, and the y-axis shows RMSD to an alternative, structurally similar TS. Displayed are the initial React-OT prediction, the fine-tuned sample, and both TS structures. 
\textbf{d} Improvement rate (left y-axis in blue) and mean reduction in energy error (right y-axis in grey) as a function of the initial React-OT RMSD.
}\label{fig:failure}
\end{figure*}

To further investigate the performance of \aefm, we conduct a detailed analysis across diverse scenarios, aiming to better understand the factors influencing its strengths and limitations. A first aspect we examine is the asymmetry in the distribution of barrier height errors, which is particularly evident for refined samples generated using React-OT as prior. Figure~\ref{fig:failure}a shows pre- and post-refinement energetic errors, where points below the bisecting line indicate improvement. An illustrative outlier contributing to the skewed mean is shown in Figure~\ref{fig:failure}c. For the particular reaction, we consider four TS, the reference (intended) TS, the React-OT prediction, its fine-tuned version obtained via \aefm, and an alternative TS associated with a different but structurally similar reaction. The plot illustrates the structural deviation, measured as RMSD, to the intended TS on the $y$-axis and to the alternative TS on the $x$-axis, while the marker color encodes the relative energy with respect to the intended TS. The original React-OT prediction deviates notably from the intended TS, with an RMSD of 0.632~\AA\ and an energy difference of 17.904~kcal~mol$^{-1}$. After fine-tuning, the sample shifts further away from the intended TS, reaching an RMSD of 0.793~\AA\ and a significantly larger energy difference of 120.993~kcal~mol$^{-1}$. At first glance, this might appear to be a failure of the optimization process. However, comparison with the alternative TS reveals a different picture, the fine-tuned structure is nearly identical to this other TS, exhibiting an RMSD of just 0.048~\AA\ and an energy deviation of merely 0.256~kcal~mol$^{-1}$. This behavior is explained by the initial proximity of the React-OT sample to the alternative TS, with an RMSD of 0.359~\AA\ compared to the intended TS. Since \aefm operates purely on structural refinement and is trained on perturbed TS geometries without access to reactant-product context, it interprets the input as a noisy version of the alternative TS and converges accordingly. To further analyse this effect, all React-OT samples were compared with similar other TS. To ensure that the alternative TSs are meaningfully closer to the sample, we only retain cases in which the RMSD to the alternative TS is at least 30\% lower than the RMSD to the originally intended TS. The mean RMSD is now improved by 7\% and the absolute energetic error by 5\% compared to the initial analysis of fine-tuned samples. This example highlights an essential characteristic of the approach, in the absence of explicit reaction context, \aefm fine-tunes samples toward structurally and energetically valid TSs, which may not always correspond to the originally intended reaction. Such behavior is typical for surface walking algorithms, where the target is to find any nearby viable TS given an initial guess structure\cite{banerjee_search_1985, baker_algorithm_1986, henkelman_dimer_1999}.

A key element influencing the performance of \aefm is the quality of the initial guess. Figure~\ref{fig:failure}c illustrates this by showing the percentage of energetically improved samples along the left $y$-axis, and the corresponding mean energy improvement along the right $y$-axis, both plotted against increasing RMSD thresholds applied to the initial React-OT samples. At each threshold, only those samples with an initial RMSD below the given value are included in the statistics. The results show a clear trend, with both the likelihood and magnitude of improvement being higher at lower RMSD thresholds. Specifically, for samples with RMSD below 0.2~\AA, 73\% of the reactions show an energetic improvement after fine-tuning, with a mean improvement of 0.15~kcal~mol$^{-1}$. In contrast, at higher thresholds, we have 69\% improved reactions and a mean energetic improvement of 0.06~kcal~mol$^{-1}$.

A notable feature of \aefm is its rapid and efficient training. Since the model operates on slightly perturbed TS structures, it converges as fast as 600 epochs. In contrast, React-OT involves a more complex training pipeline, consisting of 2000 epochs for training the diffusion model\cite{duan_accurate_2023}, followed by 200 additional epochs for the optimal transport loss\cite{duan_optimal_2025}. Beyond training time, \aefm also demonstrates strong data efficiency. To assess this, we evaluated the impact of training set size on fine-tuning CI-NEB xTB samples. As shown in Figure S1 in the appendix, increasing the amount of training data systematically reduces the energy error, both in terms of the mean and the median. Notably, using only 4000 training samples, less than half of the whole dataset, already achieves a 24.6\% reduction in mean absolute error in barrier height, compared to the 40\% reduction obtained using the full 9000 samples.

\subsection*{Physics-Informed loss improves chemical validity}\label{subsec2_4}

\begin{figure*}[ht]
\centering
\includegraphics[width=0.8\textwidth]{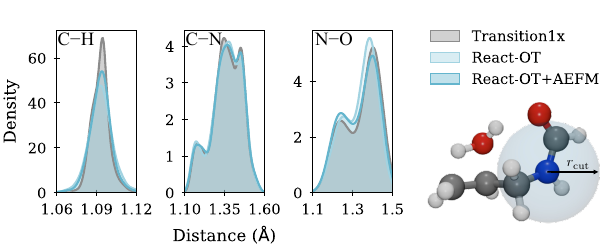}
\caption{
\textbf{Bond length distributions.}
Distributions of C–H, C–N, and N–O bond lengths in the Transition1x dataset compared to those in the React-OT and \aefm-refined structures.
}\label{fig:bond_loss}
\end{figure*}

In some cases, the potential energy surface is highly sensitive to subtle variations in bond lengths, as certain bonds contribute disproportionately to the total energy due to their stiffness. Standard coordinate-based loss functions, such as mean squared error, apply uniform weighting to atomic displacements, making them ill-suited to capture the varying energetic sensitivities across different degrees of freedom. To address this, \aefm incorporates an additional bond-based loss term. During training, the model explicitly compares the chemical environment around each atom within a 2~\AA\ cutoff radius of the ground truth TS to that of the predicted structure. For this purpose, a neighbor list is constructed for each atom in the ground truth TS by identifying all atoms within the cutoff. The same neighbor list is then applied to the predicted structure, and the corresponding interatomic distances are compared to those in the ground truth. By aligning local environments in this way, the model is encouraged to maintain realistic bonding patterns and penalize unphysical distortions, reinforcing chemical consistency and improving energetic fidelity in its predictions.

To assess the impact of the bond loss term, we compare \ aefm's fine-tuning performance when including the term versus omitting it, using two representative low-fidelity sources, React-OT\cite{duan_optimal_2025} and xTB\cite{Bannwarth2021}. For React-OT samples, incorporating the bond loss results in a 27\% reduction in the median absolute error of barrier heights. In contrast, the same model without the bond loss achieves only a 3.5\% improvement (Table~S5). To understand the source of this improvement, we analyze how the bond loss affects the model's ability to recover chemically plausible local structures. Specifically, we evaluate whether the refined structures better match the bond length distributions within a 2~\AA\ neighborhood of each atom present in the dataset. Interactions are categorized as either bonded or non-bonded based on threshold distances (Table~S3). With the bond loss, the average similarity to the reference bond length distributions improves by 35.7\% for bonded interactions and by 6\% for non-bonded ones, as illustrated for selected bonds in Figure~\ref{fig:bond_loss}. Given that bonded interactions dominate the intramolecular potential energy landscape, enhancing their accuracy is critical for reliable energy predictions. This effect is even more pronounced for xTB samples, which exhibit larger deviations from the target distribution. Here, the bond loss leads to a 57\% improvement in bonded interaction similarity and a 54\% improvement for non-bonded ones (Table~S4).

Moreover, average displacement metrics such as RMSD often fail to reflect meaningful changes in energy, underscoring their limited sensitivity, as shown in Figure~\ref{fig:stats}a and Figure~\ref{fig:failure}b. Notably, the fraction of samples that improve in both RMSD and energy is considerably smaller than the fraction that improve in energy alone (Figure~\ref{fig:stats}a). In line with this, the correlation between energetic and structural improvement is weak, with a Pearson coefficient of only 0.17 (Figure~\ref{fig:failure}b). A similarly weak relationship between RMSD and energy difference was also reported by Duan et al.\cite{duan_accurate_2023}. That highlights that generating realistic bond lengths in the refinement process is just as crucial as minimizing deviations in atomic positions. In many TS structures, the energetic accuracy is governed primarily by the reactive center. Consequently, even if the RMSD improves slightly for some atoms, introducing unrealistic bonds, such as excessively short ones, can severely degrade energetic similarity\cite{zhao2023delta}. This effect is further illustrated by the distribution of C–H bond lengths, which, after refinement with \aefm, shows a 44\% higher similarity to the dataset distribution compared to the original React-OT samples. While the refined C–H bond might not match the exact pose of the reference, its physically accurate length improves energetic similarity, even if the overall RMSD appears worse.

This observation relates to a broader challenge in molecular generative modeling, generating chemically consistent bond geometries\cite{peng_moldiff_2023,williams_physics-informed_2024, vost_improving_2025,wohlwend_boltz-1_2025, galustian_goflow_2025}. Several recent works have proposed solutions to mitigate this issue. For example, Boltz-1\cite{wohlwend_boltz-1_2025} biases generation toward low-energy configurations using physically inspired energy functions. While effective in diffusion-based generation schemes, this approach is incompatible with our fixed-point inference method, which does not rely on stochastic sampling. Vost et al.\cite{vost_improving_2025} address the sensitivity of generative models to geometric distortions by augmenting training data with perturbed structures and conditioning the diffusion process on the distortion level. However, this requires training a diffusion model from pure Gaussian noise on distortion-conditioned data, whereas our method uses an adaptive prior. Williams et al.\cite{williams_physics-informed_2024} propose a physics-informed diffusion model that decomposes the generative task into separate components for bonding, bending, torsion, and chirality, enabling more physically grounded predictions. This decomposition, however, depends on a specialized neural network architecture and limits the flexibility to choose general-purpose backbones. Finally, Falck et al.\cite{falck2025fourier} analyze the influence of the noising schedule on the recovery of high-frequency features, such as precise bond lengths. While theoretically insightful, their analysis was not conducted in the context of molecular modeling.

These efforts highlight the importance of incorporating structural or energetic priors to improve the physical fidelity of generated molecules. In contrast to more complex solutions, \aefm addresses this issue with a simple yet effective bond loss term, which guides the model toward reproducing the bond distributions found in the underlying data.

\subsection*{Quantum chemical validation}\label{subsec2_5}

While the combination of tight-binding methods or generative models with \aefm enables fast and robust high-throughput TS screening, a full quantum mechanical treatment remains essential for detailed mechanistic studies\cite{peng2016computing,von_lilienfeld_exploring_2020,unsleber_exploration_2020,nandy_computational_2021,jorner_organic_2021}. In such cases, transition states must be refined using saddle point optimization at the DFT level\cite{lam2020applications}. These optimizations typically require multiple evaluations of forces or even full Hessians, making them computationally demanding, even for small molecules\cite{koistinen_nudged_2017, yuan_analytical_2024}.

\begin{figure*}[ht]
\centering
\includegraphics[width=0.9\textwidth]{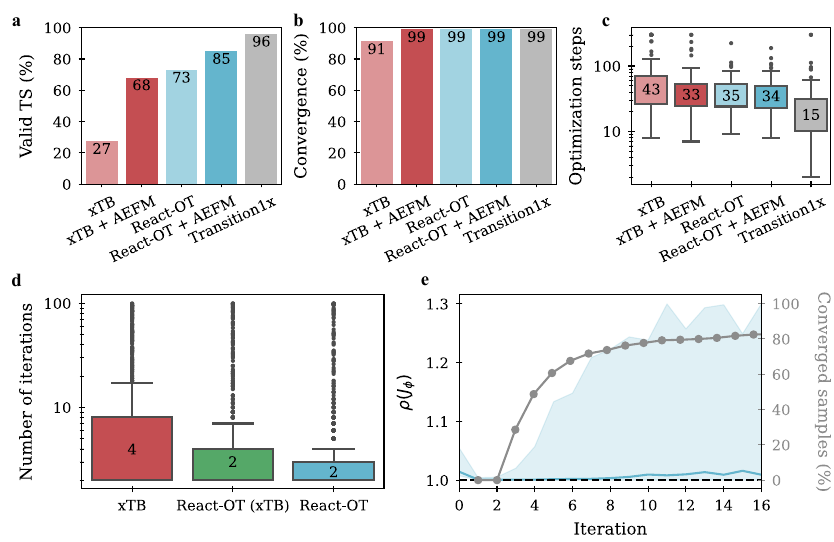}
\caption{
\textbf{Chemical validation and fixed-point convergence analysis.} 
\textbf{a} Fraction of valid TS structures, defined by the presence of exactly one imaginary frequency in the Hessian. \textbf{b} Convergence rate of DFT TS optimizations. \textbf{c} Boxplot of DFT optimization steps required to reach a converged TS structure. 
\textbf{d} Number of iterations required by \aefm to reach a fixed point. Convergence is defined by an RMSD below 0.01 between successive iterates; otherwise, inference is terminated after 100 iterations. \textbf{e} Spectral radius of the model's Jacobian with respect to the input structure, shown as median (solid line) and interquartile range (shaded region) over iterations (left y-axis). The percentage of converged samples is plotted on the right y-axis. Contractive behavior ensuring convergence occurs when $\rho(J_\phi) < 1.0$, while advanced solvers still succeed beyond this threshold.
}\label{fig:opt_conv}
\end{figure*}

To highlight the practical impact of \aefm on downstream applications, we evaluate its effect on the chemical validity of TS structures and the efficiency of DFT-based TS optimizations. For a representative set of 100 reactions, we compare three key metrics,  namely the fraction of valid TS structures (a), identified by exactly one imaginary frequency in the Hessian, the convergence rate of DFT TS optimizations (b), and the number of optimization steps required (c). Each metric is assessed for both the raw input structures and the corresponding \aefm-refined samples. \aefm incurs minimal overhead, typically requiring only 2 to 5 model evaluations depending on the quality of the initial guess, as seen in Figure~\ref{fig:opt_conv}d. In contrast, full DFT optimizations are significantly more expensive. Applied to GFN2-xTB initial guesses, \aefm increases the fraction of valid TS structures from 27\% to 68\%, a 41\% absolute improvement (Figure~\ref{fig:opt_conv}a). Moreover, \aefm improves the overall convergence rate of TS optimizations from 91\% to 99\% (Figure~\ref{fig:opt_conv}b), further underscoring its robustness. Lastly, \aefm reduces the median number of DFT optimization steps by 10, corresponding to a threefold acceleration of CPU hours needed in the refinement process (Figure~\ref{fig:opt_conv}c, Table~S2).

\subsection*{Convergence analysis}\label{subsec2_6}

As \aefm relies on fixed-point iteration, understanding its convergence behavior is critical. A standard indicator of local convergence is the Lipschitz constant $L$, which quantifies how sensitively the model output responds to input perturbations. In practice, however, this condition is often evaluated via the spectral radius $\rho(J_\phi)$, the largest absolute eigenvalue of the Jacobian $J_\phi$ at a given point $\mathbf{x}$. By Lyapunov's linearization theorem, the condition $\rho(J_\phi) < 1$ suffices for convergence in the absence of advanced solvers. However, as Bai et al.\cite{bai_stabilizing_2021} point out, this requirement can be overly conservative in practice. Methods like Broyden's method\cite{broyden1965class} or Anderson acceleration\cite{anderson1965iterative} often succeed even when $\rho(J_\phi) < 1$, due to their ability to handle mild local non-contractive behavior. To assess \ aefm's convergence characteristics, Figure~\ref{fig:opt_conv}e displays the evolution of $\rho(J_\phi)$ over refinement iterations on GFN2-xTB samples. Convergence is defined as the point where the RMSD between successive iterates falls below 0.01, as specified in Equation~\ref{eq_convergence}. If convergence is not achieved, inference is terminated after 100 iterations. The plot shows the median, along with the 25th and 75th percentiles, and overlays the cumulative convergence rate. Initially, the spectral radius drops sharply, reflecting strong local contractivity and rapid convergence. After iteration 4, the median convergence point begins to rise again. This increase does not signal failure but highlights that remaining unconverged samples tend to be more structurally complex and locally less stable. These more complicated cases dominate the later iterations, pushing the upper quantiles of $\rho(J_\phi)$ upward. Still, even in these regions, the 75th percentile remains below 1.3, indicating near-contractive dynamics. Out of 1073 React-OT and 945 xTB samples, only 6 and 3, respectively, failed to converge before reaching the iteration limit. Overall, \aefm achieves fast and stable convergence for the majority of samples, with early iterations characterized by low spectral radii and minimal computational overhead. Although convergence is slower for a few complex cases, they remain computationally manageable, with inference times not exceeding 1.6 seconds.

\section{Discussion}\label{sec3}

\aefm addresses a core challenge in reaction mechanism elucidation by converting low-fidelity TS guesses into chemically accurate, DFT-quality structures with minimal computational cost. By learning a time-independent flow field, conditioned on a prior tailored to the systematic error distribution of approximate methods, \aefm provides reliable refinements across diverse inputs. Each inference call requires only a fraction of a second per structure, enabling seamless integration into high-throughput pipelines without introducing significant computational overhead.

This makes \aefm highly suitable to enhance fast TS generators such as GFN2-xTB or React-OT, improving the chemical viability of their outputs and substantially reducing the effort required for downstream DFT saddle point optimization. This lightweight correction mechanism also opens the door to more complex applications, such as heterogeneous catalysis or enzymatic systems, where initial guesses are costly to refine and subtle structural features are often critical to reactivity.

A central strength of \aefm lies in its use of a physics-informed bond loss, which actively steers refinements toward chemically meaningful local structures, addressing a critical weakness in generative models for molecule generation. Looking ahead, incorporating higher-order geometric features, such as angles and torsions, alongside adaptive interaction cutoffs could unlock even greater accuracy and broaden applicability to more complex chemistries.

Despite its robustness, \aefm is limited by the support of its training prior. For initial guesses that deviate substantially from typical training-time errors, performance degrades. One promising path forward involves a two-stage refinement strategy guided by model uncertainty. A specialized model, trained on broader structural deviations, could be applied when the primary model signals high uncertainty, enabling robust treatment of more strongly perturbed inputs.

Finally, the principles behind \aefm generalize beyond TS refinement. In fields such as scientific machine learning, where coarse-grained simulations are used to accelerate predictions in high-dimensional systems, \aefm-like architectures could enhance the spatial and temporal resolution of neural PDE solvers. This may enable both more accurate forecasts and longer stable simulation horizons. Overall, \aefm offers a flexible and computationally efficient paradigm for lifting low-fidelity predictions to chemically and physically meaningful accuracy across a range of domains.

\section{Methods}\label{sec4}

\subsection{Flow machting}\label{subsec4_1}

Flow Matching\cite{lipman_flow_2022,liu_flow_2022} is a generative modeling approach that learns a transformation from a simple base distribution $q_0$ to a target distribution $q_1$. The base distribution $q_0$ is often referred to as the prior, and the target distribution $q_1$ as the data distribution.

To model this transformation, FM learns a continuous-time vector field \( \mathbf{v}_\theta(\mathbf{x}_t, t) \). The point \( \mathbf{x}_t \) lies along a predefined interpolation path between samples \( \mathbf{x}_0 \sim q_0 \) and \( \mathbf{x}_1 \sim q_1 \). Therefore, an optimal transport probability path\cite{tong2024improving} with the interpolation variable $t \in [0,1]$ is defined as:
\begin{equation}\label{eq_probability_path}
    p_t(\mathbf{x} \mid \mathbf{x}_0, \mathbf{x}_1) = \mathcal{N}\left(\mathbf{x} \mid (1 - t)\mathbf{x}_0 + t\mathbf{x}_1,\; \sigma_\text{FM}^2 \mathbf{I}\right),
\end{equation}
leading to samples:
\begin{equation}\label{eq_fm_lin_interpolation}
    \mathbf{x}_t = (1 - t)\mathbf{x}_0 + t\mathbf{x}_1 + \sigma_\text{FM} \boldsymbol{\epsilon}.
\end{equation}
We set \(\sigma_\text{FM}\) to 0.5 in our experiment. The corresponding target velocity field is defined as:
\begin{equation}
    \mathbf{v}_t(\mathbf{x}_t; \mathbf{x}_0, \mathbf{x}_1) = \frac{d\mathbf{x}_t}{dt} = \mathbf{x}_1 - \mathbf{x}_0.
\end{equation}
In doing so, FM models how probability mass moves over time from the prior to the data distribution. To train the vector field \( \mathbf{v}_\theta \), the squared error between the predicted velocity and the target velocity is minimized. The training objective, known as the FM loss, is given by:
\begin{equation}
    \mathcal{L}_{\mathrm{FM}} = \mathbb{E}_{\mathbf{x}_0, \mathbf{x}_1, t} 
    \left[ \left\| \mathbf{v}_\theta(\mathbf{x}_t, t) - (\mathbf{x}_1 - \mathbf{x}_0) \right\|^2 \right],
\end{equation}
As an alternative loss formulation, the model $\phi_{\theta}(\mathbf{x}_t, t)$ can be trained to directly predict $\mathbf{x}_1$ at time $t$ instead of the velocity, a strategy that has demonstrated improved performance in practice\cite{stark_harmonic_2024}. This approach is commonly referred to as variational flow matching (VFM)\cite{amini2024variational}. Once $\phi_{\theta}(\mathbf{x}_t, t)$ is trained, new samples can be generated by solving the following ordinary differential equation (ODE) forward in time:
\begin{equation}\label{eq_ode}
    \frac{d\mathbf{x}_t}{dt} = \frac{\phi_{\theta}(\mathbf{x}_t, t)-\mathbf{x}_t}{1-t}, \quad \mathbf{x}_0 \sim q_0.
\end{equation}
This integration starts from a sample \( \mathbf{x}_0 \) drawn from the prior \( q_0 \), and produces a sample \( \mathbf{x}_1 \sim q_1 \) at time \( t = 1 \), using any black-box ODE solver. Later, we will make $\phi_{\theta}(\mathbf{x}_t, t)$ time-independent and use it to iteratively refine approximate solutions $\mathbf{x}^{k+1} = \phi_{\theta}(\mathbf{x}^k)$.

\subsection{Details about \aefm}\label{subsec4_2}

\subsubsection{Training}\label{subsubsec4_2_1}

A central component of \aefm is its adaptive behavior, which arises from the formulation of the source distribution $p_0$ that we learn to map to the target distribution $p_1$. In our case, the target distribution is determined by the high-fidelity TSs from the Transition1x dataset\cite{schreiner_transition1x_2022}. Given a sample $\mathbf{x}_1 \sim p_1$, we define the corresponding source sample $\mathbf{x}_0 \sim p_0$ as a noisy perturbation of $\mathbf{x}_1$:
\begin{equation}\label{eq_source_definition}
    \mathbf{x}_0 = \mathbf{x}_1 + \sigma \boldsymbol{\epsilon}, \quad \boldsymbol{\epsilon} \sim \mathcal{N}(0, \mathbf{I}).
\end{equation}
The key parameter in this formulation is $\sigma$, which controls the extent to which the source distribution deviates from the target. We assume that the deviation of low-fidelity samples $\mathbf{x}_1^\text{w}$ from their corresponding high-fidelity TSs can be modeled as Gaussian noise. %Under this assumption, our objective is to select $\sigma$ such that the resulting prior distribution approximates the low-fidelity distribution in expectation.

Under this assumption, we want to determine the noise scale $\sigma$ such that the expected error from a Gaussian corruption process with variance $\sigma^2$ matches the expected error between low-fidelity and reference TSs. Specifically, we set
\begin{align}
     \mathop{\mathbb{E}}\limits_{\substack{
        \mathbf{x}_1 \sim D, \\
        \mathbf{x}_0 \sim \mathcal{N}(\mathbf{x}_1, \sigma^2I)
    }}\left[ \frac{\|\mathbf{x}_0 - \mathbf{x}_1\|^2}{N(\mathbf{x}_1)} \right] 
    = \mathop{\mathbb{E}}\limits_{\substack{
        (\mathbf{x}_1, \mathbf{x}_1^\text{w}) \sim D
    }} \left[\frac{\|\mathbf{x}_1^\text{w} - \mathbf{x}_1\|^2 }{N(\mathbf{x}_1)} \right] ,
\end{align}
where \(N(\mathbf{x}_1)\) is the number of atoms involved. Since \(\mathbf{x}_0 = \mathbf{x}_1 - \sigma \boldsymbol{\epsilon}\), the left-hand side simplifies to
\begin{equation}
\mathbb{E}_{\boldsymbol{\epsilon}} \left[ \frac{\left\| \sigma \boldsymbol{\epsilon} \right\|^2 }{N(\boldsymbol{\epsilon})} \right]
=  \frac{\sigma^2}{N(\boldsymbol{\epsilon})} \cdot 3N(\boldsymbol{\epsilon}) = 3\sigma^2.
\end{equation}
Solving for \(\sigma\), we obtain
\begin{equation}\label{eq:sigma_definition}
\sigma =  \left(\mathbb{E}_{(\mathbf{x}_1, \mathbf{x}_1^\text{w}) \sim D} \left[ 
    \frac{ \left\| \mathbf{x}_1 - \mathbf{x}_1^\text{w} \right\|^2}{3N(\mathbf{x}_1)}
\right] \right) ^{1/2}.
\end{equation}
Thus, $\sigma$ can be calculated using the mean RMSD of the low-fidelity samples. The source distribution in our setup is designed to model the expected deviation of the low-fidelity predictions from the reference TS. It captures the distribution of typical errors observed in the low-fidelity method and provides a learning signal during training. However, in contrast to the standard FM framework, we do not sample from the prior during inference. Instead, we start from the actual output of the low-fidelity model. As a result, the model learns from the prior during training, but at inference time, it needs to adapt to the specific error of each low-fidelity input. These errors can vary considerably, with some samples being very close to the true TS and others deviating more. Assigning a uniform time value of $t=0$ to all such samples during inference, as done in conventional FM, may lead to over- or under-correction by the model. To address this, we remove explicit time conditioning during training, allowing the model to implicitly infer the quality of a given input $\mathbf{x}_t$. This helps the model estimate how far each sample is from the final prediction target. In practice, this behavior is encouraged through the use of a direct $\mathbf{x}_1$-prediction loss, as described earlier, while omitting time as an input to the network.
\begin{equation}\label{eq_loss_AEFM}
    \mathcal{L_\text{\aefm}} = \mathbb{E}_{\mathbf{x}_0, \mathbf{x}_1, t}\left[\|\mathbf{x}_1 - \phi_{\theta}(\mathbf{x}_t)\|^2 \right].
\end{equation}

\subsubsection{Inference}\label{subsubsec4_2_2}

Since we omit the concept of time, we no longer integrate the ODE from Equation~\ref{eq_ode}. Instead, we train a neural network $\phi_{\theta}$ to directly predict the endpoint $\mathbf{x}_1$ of a dynamical process starting from an initial point $\mathbf{x}_0$. This formulation aligns with the perspective of VFM, where learning a velocity field that matches trajectories between $\mathbf{x}_0$ and $\mathbf{x}_1$ can be reinterpreted as minimizing a divergence between model and reference endpoint distributions. In our case, although we do not instantiate or evaluate $\mathbf{v}_\theta(\mathbf{x}_t, t)$ directly at test time, the network’s prediction implicitly corresponds to the result of integrating such a field over time. In this sense, our model acts as a learned approximation of the ODE solution operator. To further refine predictions and ensure consistency with underlying dynamics, we employ a fixed-point iteration scheme at inference time:
\begin{equation}\label{eq_fixed_point_definition}
\mathbf{x}^{k+1} = \phi_{\theta}(\mathbf{x}^k),
\end{equation}
where the initial guess $\mathbf{x}^0$ is taken as the low-fidelity prediction, $\mathbf{x}_1^\text{w}$. Conceptually, this mirrors the inference procedure in Deep Equilibrium Models\cite{bai_deep_2019,bai_stabilizing_2021}, where a neural network is iterated to convergence at test time to find a fixed point $\mathbf{x}^*$ satisfying $\mathbf{x}^* = f_\theta(\mathbf{x}^*)$. To perform the iteration, one may employ any fixed-point solver, such as Broyden’s method\cite{broyden1965class} or Anderson acceleration\cite{anderson1965iterative}. In this work, we use the latter, which enhances convergence by leveraging multiple previous iterates and their residuals to extrapolate a more accurate fixed point. Given $m$ previous iterates $\mathbf{x}^{k-m}, \ldots, \mathbf{x}^k$ and corresponding residuals $\mathbf{g}(\mathbf{x}^i)=\phi_{\theta}(\mathbf{x}^i) - \mathbf{x}^i$, the method solves a least-squares problem to find coefficients $\boldsymbol{\alpha}$ such that the weighted sum of residuals $\sum_{i=0}^{m} \alpha_i \mathbf{g}(\mathbf{x}^{k-m+i})$ is minizimed. Given $\boldsymbol{\alpha}$, the next iterate is computed as:
\begin{equation}\label{eq_fixed_point_anderson_acceleration}
\mathbf{x}^{k+1} = \beta \sum_{i=0}^{m} \alpha_i \phi_{\theta}(\mathbf{x}^{k-m+i}) + (1 - \beta) \sum_{i=0}^{m} \alpha_i \mathbf{x}^{k-m+i},
\end{equation}
where $ \beta \in [0,1] $ is a damping parameter and $ \sum_i \alpha_i = 1 $. The fixed-point iteration is terminated once the RMSD between successive iterates falls at or below a threshold of 0.01: 
\begin{equation}\label{eq_convergence}
  \frac{\|\mathbf{x}^{k+1}-\mathbf{x}^{k}\|}{N(\mathbf{x}_k)} \leq 0.01  
\end{equation}
If the convergence criterion is not satisfied, inference is terminated after a maximum of 100 iterations. The damping parameter $\beta$ is set to 1.0 and the history size $m$ to 5, based on a hyperparameter search.

\subsection{Physical consistency loss}\label{subsec4_3}

To address issues such as bond length inconsistencies and atomic clashes in generative models, we introduce an additional loss term focused on bonding\cite{peng_moldiff_2023,williams_physics-informed_2024,wohlwend_boltz-1_2025,vost_improving_2025,galustian_goflow_2025}. This is particularly important because the PES is highly sensitive to small geometric deviations. In some cases, accurately reproducing critical bond lengths is more important than minimizing the overall positional error. A prediction may yield a low RMSD while still introducing small but chemically significant distortions in key bonds, resulting in large energetic errors. To improve the chemical plausibility of generated structures, we compare the local environment of each atom within a cutoff radius $r_\text{cut}$ to that of the corresponding atom in the ground truth structure, as shown in Figure~\ref{fig:bond_loss}.
\begin{multline}
\mathcal{L}_\text{b} = 
\mathop{\mathbb{E}}
\Bigg[ 
  \sum_{(i,j) \in \mathcal{B}(\mathbf{x}_1)}  \frac{\left[
     d_{ij}(\phi_{\theta}(\mathbf{x}_t)) - d_{ij}(\mathbf{x}_1)
  \right]^2}{\left| \mathcal{B}(\mathbf{x}_1) \right|} 
\Bigg]
\label{eq_loss_bond_definition}
\end{multline}
\begin{equation}
\mathcal{B}(\mathbf{x}_1) := \left\{ (i,j) \,\middle|\, \left\| \mathbf{x}_{1,i} - \mathbf{x}_{1,j} \right\| < r_\text{cut} \right\}
\label{eq_loss_bond_mask}
\end{equation}
with $d_{ij}=\| \mathbf{x}_{i} - \mathbf{x}_{j} \|$ as the euclidian distance between atom $i$ and $j$. The cutoff radius is set to 2~$\text{\AA}$, based on the longest equilibrium bond lengths typically observed in C, N, O, and H chemistry, with an added margin to accommodate extended bond distances that may arise in transition state structures\cite{rdomingo_new_2014}. Thus, the total loss used in training is:
\begin{equation}\label{eq_loss_total}
    \mathcal{L} = \mathcal{L_\text{\aefm}} + w_\text{b}\mathcal{L}_\text{b}
\end{equation}
with $w_\text{b}$ as a hyperparameter to weight the bond loss influence during training, which we fix to 1.0.

\subsection{Chemical validation}\label{subsec4_4}

To compute the electronic energy of samples, we use ORCA5.0.4\cite{ORCA5} in combination with ASE\cite{ase-paper} at the same level of theory as the transition1x dataset\cite{schreiner_transition1x_2022} was generated with $\omega$B97x/6-31G(d)\cite{ ditchfield1971self, chai2008systematic}. To generate the GFN2-xTB\cite{Bannwarth2021} TS guesses, CI-NEB\cite{henkelman_climbing_2000} using ASE and the python interface tblite. For the CI-NEB computations, the same protocol is used as for \textit{transition1x} generation. The NEB calculation is first run until the maximum force perpendicular to the path falls below a threshold of 0.5~eV$\text{\AA}^{-1}$. Subsequently, the CI-NEB refinement continues until convergence, defined as a maximum perpendicular force below 0.05~eV$\text{\AA}^{-1}$ or a maximum of 500 iterations. Reactions that do not meet this criterion are considered not converged. For TS optimization, the Sella package\cite{hermes2022} using the P-RFO\cite{banerjee_search_1985,baker_algorithm_1986} algorithm along with the ASE ORCA calculator is run until the maximum force of 0.001 eV$\text{\AA}^{-1}$ is achieved with a maximum number of 300 iterations. Numerical Hessians are computed using finite central difference method with an $\delta$ of 0.01$\text{\AA}$.

\subsection{Metrics}\label{subsec4_5}

The RMSD for molecules is determined by first aligning the molecules $\mathbf{x}_1$ and $\mathbf{x}_2$ using the Kabsch algorithm and then computing:
\begin{equation}\label{eq_rmsd_definition}
\begin{split}
    \text{RMSD}(\mathbf{x}_1,\mathbf{x}_2) ={}& 
    \sqrt{\frac{\sum_{i=1}^{N} \|\mathbf{x}_{1,i} - \mathbf{x}_{2,i} \|^2}{N}} \\
    ={}& \sqrt{\frac{\sum_{i=1}^{N} \sum_{j \in \{x,y,z\}} (x_{1,i,j} - x_{2,i,j})^2}{N}}
\end{split}
\end{equation}
with $N$ denoting the number of atoms. Note that this definition differs from the one used in React-OT, where the RMSD is normalized by $3N$ instead. To access the difference in barrier height, the electronic energy $V$ of each sample TS structure is computed and the MAE is defined as
\begin{equation}\label{eq_mae_definition}
    \text{MAE}= \frac{1}{M} \sum_{i}^{M} |V(\mathbf{x}_i) - V(\hat{\mathbf{x}}_i)|
\end{equation}
with $\hat{\mathbf{x}}_i$ as the predicted TS and $\mathbf{x}_i$ as the corresponding database TS and $M$ as the total number of samples. To compare the distribution of bond lengths in the predicted structures with those in the reference data, we use the Wasserstein-1 distance. Given two one-dimensional empirical distributions $p$ and $q$ over bond lengths with cumulative distribution functions $P$ and $Q$, respectively, the Wasserstein-1 distance is defined as:
\begin{equation}
    W_1(p, q) = \int_{-\infty}^{\infty} \left| P(x) - Q(x) \right| \, dx.
\end{equation}
The Wasserstein-1 distance is computed separately for each bond type in the dataset and subsequently averaged across all types. As a metric that quantifies the minimal effort required to transform one distribution into another, it is particularly well-suited for capturing differences in geometric structure distributions, such as bond lengths.

\section*{Data Availability}
The Transition1x dataset\cite{schreiner_transition1x_2022} used in this work is available via Figshare at https://doi.org/10.6084/m9.figshare.19614657.v4. 
% The pretrained AEFM model checkpoints and the corresponding databases for each low-fidelity source are available on Zenodo at \textcolor{red}{[DOI/URL to be inserted]}.

\section*{Code Availability}
% The AEFM codebase is publicly available as an open-source repository on GitHub to support continuous development. A stable release version has been archived on Zenodo at \textcolor{red}{[DOI/URL to be inserted]}.
The code is currently under review and will be available as an open-source repository on GitHub.

\section*{Acknowledgements}
This research was funded by the Ministry of Science, Research and the Arts Baden-Wuerttemberg in the Artificial Intelligence Software Academy (AISA). We thank the Deutsche Forschungsgemeinschaft (DFG, German Research Foundation) for supporting this work by funding - EXC2075 – 390740016 under Germany's Excellence Strategy. Tanja was supported by the Deutsche Forschungsgemeinschaft (DFG, German Research Foundation) under Germany’s Excellence Strategy – EXC 2120/1 – 390831618. We acknowledge the support by the Stuttgart Center for Simulation Science (SimTech). The authors acknowledge support by the state of Baden-Württemberg through bwHPC and the German Research Foundation (DFG) through grant no INST 40/575-1 FUGG (JUSTUS 2 cluster). We thank Moritz, Anji, and others for insightful discussions throughout the development of this work. We are also grateful to Boshra for her valuable input on improving the visual presentation of the manuscript.

\section*{Author Contributions}

S.D. led the project, contributing to the conceptualization, methodology, software development, validation, investigation, data curation, drafting of the manuscript, as well as its review, editing, and visualization. V.T. supported the methodology and software development and contributed to manuscript editing. T.B. was involved in methodology, visualization, and manuscript review. J.K. provided supervision, contributed to manuscript review and editing, and secured funding. M.N. contributed to the conceptualization, methodology, and visualization, co-wrote the initial draft, and played a central role in manuscript review, editing, supervision, and funding acquisition.

\section*{Competing Interests}
The authors declare no competing interests.

% \bibliographystyle{naturemag}
% \bibliography{final_literature}

\end{document}

\raggedbottom
\maketitle

\section{Comparison to flow matching}

An alternative to \aefm is to apply flow matching (FM) using a Gaussian prior and the transition states as the target distribution. The initial time from which the ODE in FM is integrated is inferred based on the mean RMSD between the low-fidelity structures and the reference transition states, using the definition of the intermediate interpolants $\mathbf{x}_t$. Specifically, $t_0$ is chosen such that the RMSD between the time interpolant $\mathbf{x}_t$ and the target $\mathbf{x}_1$, $\nicefrac{\|\mathbf{x}_t - \mathbf{x}_1\|}{N}$, matches the average RMSD of the low-fidelity source. For GFN2-xTB samples, this yields $t_0 = 0.87$, and for React-OT samples, $t_0 = 0.93$. Table~\ref{tab:performance_fm} reports the corresponding performance. These results are significantly worse than those obtained with \aefm, which can be attributed to the fact that FM must learn the flow field from a Gaussian prior, making the task considerably more complex compared to \aefm. Furthermore, to ensure a fair comparison with \aefm, the source and target molecules are not aligned, resulting in a non-linear vector field that is harder to integrate and leads to less accurate structures.

\begin{table}[ht]
\centering
\caption{Performance using FM for refinement.}\label{tab:performance_fm}
\begin{tabular}{@{\extracolsep\fill}lllll}
\toprule
Approach & \multicolumn{2}{c}{RMSD (\AA)} & \multicolumn{2}{c}{$|\Delta E_{\text{TS}}|$ (kcal mol$^{-1}$)} \\
\cmidrule(lr){2-3} \cmidrule(lr){4-5}
& Mean & Median & Mean & Median \\
\midrule
xTB CI-NEB  & 0.312  & 0.179 & 10.426 & 2.673 \\
\rowcolor{lightblue}xTB CI-NEB + FM & 0.439  & 0.310 & 91.732 & 88.953  \\
React-OT  & 0.183  & 0.092 & 3.405  & 1.092  \\
\rowcolor{lightblue}React-OT + FM & 0.252 & 0.167 & 34.723 & 34.265 \\
\bottomrule
\end{tabular}
\end{table}

\section{Ablation and model architecture}

To evaluate the data efficiency of \aefm, the model was trained using subsets of 2000, 4000, 6000, 8000, and all 9000 training samples. Each trained model was then applied to the GFN2-xTB samples, and the resulting mean and median energetic differences to the ground truth transition states are compared in Figure~\ref{fig:data_ablation}. The results reveal a clear trend of decreasing energetic difference with increasing training data, with 4000 samples already providing a substantial improvement in mean energetic difference. Table~\ref{tab:wasserstein_distances} shows the Wasserstein-1 distance for each bonded interaction using the thresholds defined in Table~\ref{tab:bond_thresholds}. Combining React-OT with \aefm results in consistently lower Wasserstein-1 distances across nearly all bonded and non-bonded interactions, indicating improved agreement with the underlying Transition1x dataset. The model and training parameters are shown in Table~\ref{tab:hyp} and Table~\ref{tab:training}.

\begin{table}
    \centering
    \begin{minipage}[t]{0.48\textwidth}
       \vspace{0pt} % align tops properly
        \centering
        \includegraphics[width=\linewidth]{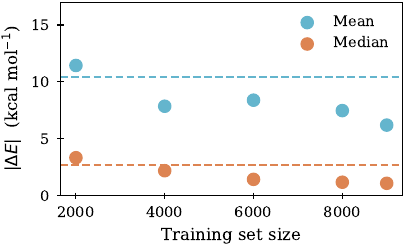}
        \captionof{figure}{
            \textbf{Data efficiency of \aefm.}  
            Mean and median energy errors for models trained with different sample sizes. Dashed lines show corresponding errors from GFN2-xTB.
        }
        \label{fig:data_ablation}
    \end{minipage}
    \hfill
    \begin{minipage}[t]{0.48\textwidth}
       \vspace{0pt} % align tops properly
        \centering
        \captionof{table}{Total CPU hours required for TS refinement using the p-RFO algorithm implemented in the Sella package with 48 CPU cores.}
        \begin{tabular}{lr}
            \toprule
            \textbf{Method} & \textbf{CPU Hours} \\
            \midrule
            xTB CI-NEB & 1430 \\
            \rowcolor{lightblue}xTB CI-NEB + \aefm & 506  \\ 
            React-OT & 455  \\
            \rowcolor{lightblue}React-OT + \aefm & 439  \\
            \bottomrule
        \end{tabular}
    \end{minipage}
\end{table}

\begin{table}
\centering
\caption{Threshold distances used to determine bonded atom pairs. To categorize into bonded and non-bonded, an additional margin of 0.1\,\AA\ is added on top of the threshold values.}
\label{tab:bond_thresholds}
\begin{tabular}{lccccccccc}
    \toprule
     & C--C & C--H & C--N & C--O & H--H & H--N & H--O & N--N & N--O \\
    \midrule
    Bond Threshold (\AA) & 1.54 & 1.09 & 1.47 & 1.43 & 0.74 & 1.01 & 0.96 & 1.45 & 1.40 \\
    \bottomrule
\end{tabular}
\end{table}

\begin{table}
\centering
\caption{Wasserstein-1 distance to bond distribution inherent in the test samples of the Transition1x dataset for different bonded (bd) and non-bonded (nbd) atom pairs (lower is better).}
\label{tab:wasserstein_distances}
\begin{tabular}{@{\extracolsep{\fill}}lccccccccc}
\toprule
Method & C--C & C--H & C--N & C--O & H--H & H--N & H--O & N--N & N--O \\
\midrule
xTB CI-NEB (bd) & 0.0041 & 0.0058 & 0.0097 & 0.0065 & 0.0151 & 0.0046 & 0.0029 & 0.0085 & 0.0113 \\
\rowcolor{lightblue}xTB CI-NEB + \aefm (bd) & 0.0032 & 0.0015 & 0.0021 & 0.0018 & 0.0543 & 0.0063 & 0.0140 & 0.0116 & 0.0099 \\
React-OT (bd)  & 0.0022 & 0.0019 & 0.0033 & 0.0019 & 0.0124 & 0.0035 & 0.0059 & 0.0080 & 0.0142 \\
\rowcolor{lightblue}React-OT + \aefm (bd)  & 0.0011 & 0.0011 & 0.0023 & 0.0014 & 0.0081 & 0.0034 & 0.0061 & 0.0059 & 0.0110 \\
\midrule
xTB CI-NEB (nbd) & 0.0067 & 0.0289 & 0.0192 & 0.0129 & 0.0082 & 0.0461 & 0.0913 & -- & 0.0697 \\
\rowcolor{lightblue}xTB CI-NEB + \aefm (nbd) & 0.0046 & 0.0097 & 0.0138 & 0.0234 & 0.0027 & 0.0324 & 0.0416 & 0.0981 & 0.0466 \\
React-OT (nbd) & 0.0074 & 0.0068 & 0.0128 & 0.0115 & 0.0037 & 0.0088 & 0.0205 & 0.1957 & 0.0413 \\
\rowcolor{lightblue}React-OT + \aefm (nbd) & 0.0061 & 0.0065 & 0.0106 & 0.0122 & 0.0034 & 0.0139 & 0.0184 & 0.0851 & 0.0518 \\
\bottomrule
\end{tabular}
\end{table}

\begin{table}
\centering
\caption{Performance using \aefm without bond loss for refinement.}\label{tab:performance_aefm}
\begin{tabular}{@{\extracolsep\fill}lllll}
\toprule
Approach & \multicolumn{2}{c}{RMSD (\AA)} & \multicolumn{2}{c}{$|\Delta E_{\text{TS}}|$ (kcal mol$^{-1}$)} \\
\cmidrule(lr){2-3} \cmidrule(lr){4-5}
& Mean & Median & Mean & Median \\
\midrule
xTB CI-NEB  & 0.312  & 0.179 & 10.426 & 2.673 \\
\rowcolor{lightblue}xTB CI-NEB + \aefm ($w_\text{b}$=0.0) & 0.249  & 0.103 & 10.405 & 1.518  \\
React-OT  & 0.183  & 0.092 & 3.405  & 1.092  \\
\rowcolor{lightblue}React-OT + \aefm ($w_\text{b}$=0.0)& 0.186 & 0.083 & 3.750 & 1.056 \\
\bottomrule
\end{tabular}
\end{table}

\begin{table}
\centering
\begin{minipage}[t]{0.48\linewidth}
\centering
\caption{Model hyperparameters.}\label{tab:hyp}
\begin{tabular}{@{\extracolsep\fill}ll}
\toprule
Parameter & Value \\
\midrule
Message passing layers & 6 \\
Equivariant readout layers & 1 \\
Hidden features & 196 \\
Radial basis functions & 96 \\
Cutoff radius & 10\,\AA \\
\bottomrule
\end{tabular}
\end{minipage}
\hfill
\begin{minipage}[t]{0.48\linewidth}
\centering
\caption{Training hyperparameters for low-fidelity sources.}\label{tab:training}
\begin{tabular}{@{\extracolsep\fill}lcc}
\toprule
Method & $\sigma$ & Epochs \\
\midrule
xTB CI-NEB & 0.19 & 1000 \\
React-OT (xTB) & 0.12 & 600 \\
React-OT & 0.11 & 600 \\
\bottomrule
\end{tabular}
\end{minipage}
\label{tab:side_by_side_tables}
\end{table}